 \documentclass[preprint2]{aastex} 

\slugcomment{}

\shorttitle{PThe relationships of solar flares with both sunspot
and geomagnetic activity}
  \shortauthors{Z. L. Du \& H. N. Wang}

\begin{document}
\title{The relationships of solar flares with both sunspot and geomagnetic
activity}

\author{Z. L. Du and H. N. Wang\altaffilmark{}}
\affil{Key Laboratory of Solar Activity, National Astronomical
Observatories, Chinese Academy of Sciences, Beijing 100012, China}
\email{zldu@nao.cas.cn}

\begin{abstract}
The relationships between solar flare parameters (total
importance, time duration, flare index, and flux) and sunspot
activity ($R_\mathrm{z}$) as well as those between geomagnetic
activity ($aa$ index) and the flare parameters can be well
described by an integral response model with the response time
scales of about eight and thirteen months, respectively. Compared
with linear relationships, the correlation coefficients of the
flare parameters with $R_\mathrm{z}$, of $aa$ with the flare
parameters, and of $aa$ with $R_\mathrm{z}$ based on this model
have increased about 6\%, 17\%, and 47\% on average, respectively.
The time delays between the flare parameters with respect to
$R_\mathrm{z}$, $aa$ to the flare parameters, and $aa$ to
$R_\mathrm{z}$ at their peaks in solar cycle can be predicted in
part by this model (82\%, 47\%, and 78\%, respectively). These
results may be further improved when using a cosine filter with a
wider window. It implies that solar flares are related to the
accumulation of solar magnetic energy in the past through a time
decay factor. The above results may help to understand the
mechanism of solar flares and to improve the solar flare
prediction.
\end{abstract}

\section{Introduction}           
     \label{sect:intr}

Solar flares are powerful eruptions of solar activity
\citep{Ozguc89,Mikic94,Jain10,Fang11} occurring on time scales of
minutes up to a few hours \citep{Chandra11} and may produce a
series of solar-terrestrial effects, which may be hazardous to
both spacecraft and astronauts. Understanding the mechanism of
solar flares and forecasting them are important for both solar
physics and geophysics. Several mechanisms have been proposed to
explain the eruptions of solar flares, such as the photospheric
converging and shear motions \citep{Mikic94}, flux emergence and
cancellation \citep{Gan93,Zhang01}, catastrophe model
\citep{Forbes90}, and Kink instability of coronal flux ropes
\citep{Sakurai76,LiGan11}. The magnetic reconnection plays an
important role in triggering solar flares
\citep{Lin95,Wheatland01,Forbes06,Fang10}.

To quantify the daily flare activity over 24 hours per day,
\citet{Kleczek52} introduced the `flare index' defined as
\begin{equation}
  \label{Eq:q}
        Q=i\times t,
\end{equation}
where `$i$' represents the intensity scale of importance and `$t$'
the duration (in minutes) of the flare \citep{Knoska84,Atac98}.
This relationship is assumed to give roughly the total energy
emitted by a flare \citep{Kleczek52}. The solar flare activity is
found to be closely correlated with sunspots
\citep{Ozguc89,Feminella97}. Larger flares appear often near
larger and more complex active regions
\citep{McIntosh90,Bachmann94,Norquist11}. Sunspot activity is a
striking manifestation of magnetic fields on the Sun, associated
with the main sites of solar-activity phenomena \citep{Moradi10}
and related to the energy supplied into the corona
\citep{Toma00,Temmer03}. Studying the relationship between solar
flares and sunspot activity is useful to understand and predict
the former. The flare frequency of occurrence is often predicted
by sunspot groups or numbers
\citep{McIntosh90,Gallagher02,Cui06,Yu09,Huang10} for increasing
applications in space weather.

Solar activity is well known to be at the origin of geomagnetic
activity \citep{Snyder63,Crooker77}. Studying the relationship
between solar activity, as represented by the International
sunspot number ($R_\mathrm{z}$), and geomagnetic activity, as
represented by the $aa$ index \citep{Mayaud72}, is useful for
understanding the formation of the latter and the mechanism of
solar cycle \citep{Feynman78,Legrand89,Du11a,Du210,Du211a,Du211b}.
Conventionally, the relationship between $aa$ and $R_\mathrm{z}$
is often analyzed by point-point correspondence. However, some
questions are hardly understood such as the significant increase
in the $aa$ index over the twentieth century
\citep{Feynman78,Clilverd98,Lukianova09}, and the variations in
the correlation between $aa$ and $R_\mathrm{z}$
\citep{Borello92,Echer04,Du11b}. It is found that these phenomena
can be well explained by an integral response model recently
presented by \citet{Du11c}. The value of $aa$ depends not only on
the present $R_\mathrm{z}$ but also on past values.

The geomagnetic activity results from various phenomena which are
related to the interplanetary magnetic field
\citep[IMF,][]{Stamper99}, solar wind
\citep{Svalgaard77,Legrand89,Tsurutani95}, Coronal Mass Ejection
\citep[CME,][]{Legrand89}, galactic cosmic rays \citep{Stamper99},
and others \citep{Legrand89,Stamper99}. \citet{Gosling93} pointed
out that CMEs, rather than flares, were the critical element for
large geomagnetic storms, interplanetary shocks, and major solar
energetic particle (SEP) events, which was argued by
\citet{Richardson02}.

This study analyzes the relationships between solar flare
parameters (Section~\ref{sec:Data}) and $R_\mathrm{z}$ as well as
the relationships between the $aa$ index and the flare parameters
using an integral response model \citep{Du11c} in
Sections~\ref{subsec:I}-\ref{subsec:F}. Conclusions are summarized
finally in Section~\ref{sec:Discussions}.  

\section{Data}
\label{sec:Data}

The data used are the time series of monthly mean geomagnetic $aa$
index\footnote{ftp://ftp.ngdc.noaa.gov/STP/SOLAR\_DATA/RELATED\-\_INDICES/AA\_INDEX/}
\citep{Mayaud72}, the international sunspot number
($R_{\mathrm{z}}$)\footnote{http://www.ngdc.noaa.gov/stp/spaceweather.html},
and solar flare parameters based on Geostationary Operational
Environmental Satellite (GOES) soft X-ray flares shown as
follows\footnote{ftp://ftp.ngdc.noaa.gov/STP/SOLAR\_DATA/}.
\begin{enumerate}
  \item[(i)] $I$: total importance of flares, $I= 100X+10M+C+0.1B$, where $X$, $M$, $C$, and $B$
  are the flare classes  \citep{Cui06}.
  \item[(ii)] $T$: time duration of flare (in minutes) .
  \item[(iii)] $Q$: the `flare index' from Equation (\ref{Eq:q}) by \citet{Kleczek52}.
  \item[(iv)] $F$: flux from event start to end (in J/m$^2$) .
\end{enumerate}

 \begin{figure}[!tb]
 \includegraphics[width=\columnwidth]{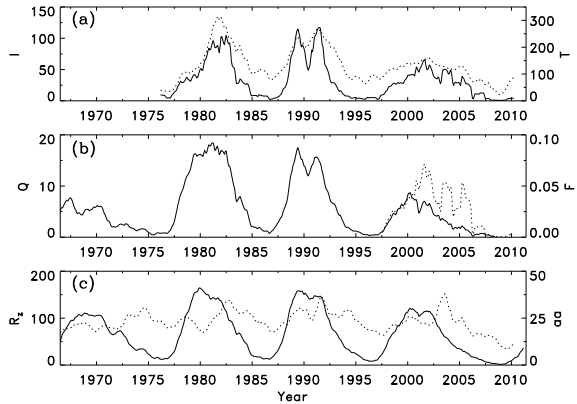}
 \caption{ (a) $I$ (solid) and $T$ (dotted) since March 1976, with a correlation coefficient of $r=0.89$.
     (b) $Q$ (solid) since July 1966 and $F$ (dotted) since July 1997, with a correlation coefficient of $r=0.67$.
     (c) $R_\mathrm{z}$ (solid) and $aa$ (dotted) since July 1966, with a correlation coefficient of $r=0.40$.
     }
 \label{Fig:1}
 \end{figure}

These parameters are first summed over each day and, then,
averaged over each month to obtain the monthly means of the daily
integrated quantities. To filter out high frequency variations in
the data, the parameters are smoothed with the commonly used
13-month running mean technique. The solar flare parameters since
July 1966 are shown in Fig.~\ref{Fig:1}(a) and (b). It is seen
that these parameters are well correlated. For example, $I$ is
well correlated with $T$ ($r=0.89$, Fig.~\ref{Fig:1}(a)), and $Q$
is well correlated with $F$ ($r=0.67$, Fig.~\ref{Fig:1}(b)), both
being significant at the 99\% level of confidence.
Figure~\ref{Fig:1}(c) depicts the time series of $R_\mathrm{z}$
(solid) and $aa$ (dotted) with a correlation coefficient of
$r=0.40$.

\section{Results} \label{sec:Results}

It is well known that solar flares tend to lag behind
sunspot activity by several 
months \citep{Wheatland01,Temmer03} or even a few years
\citep{Wagner88,Aschwanden94}. To have a better understanding of
the relationships and time delays between solar flares and
$R_\mathrm{z}$, we employ the following integral response
model~\citep{Du11c} to study the relationships between the flare
parameters ($P=I$, $T$, $Q$ and $F$) and $R_\mathrm{z}$,
\begin{eqnarray}
    \label{Eq:model1}
   \begin{array}{lrl}
    y(t)     &=& D\int_{-\infty}^t x(t') e^{-(t-t')/\tau}dt'+y_0 \\
    &=& D\sum_{t'=t_0}^t x(t') e^{-(t-t')/\tau}+y_0, \\
   \end{array}
\end{eqnarray}
where $y_0$ is a constant, reflecting the part of $y=P$ that is
uncorrelated to $x=R_\mathrm{z}$ (related to other phenomena); $D$
is the `dynamic response factor' of $y$ to $x$, representing the
initial generation efficiency of $y$ by $x$ ($\partial y/\partial
x|_{t'=t}$); and $\tau$ is the `response time scale' of $y$ to
$x$, indicating the dependence of the current $y(t)$ on the past
$x(t')$ through a time decay factor $e^{-(t-t')/\tau}$ ($\tau=0$
reflects the point-point correspondence of $y$ to $x$, i.e., the
current $y(t)$ is only related to the current $x(t)$;
$\tau=+\infty$ represents that $y$ is uncorrelated to $x$). In
application, both $y$ and $x$ are discrete variables. Therefore,
we use the second formula in Equation~(\ref{Eq:model1}) with the
summation being taken over from the starting time ($t_0$) of the
series (see Fig.~\ref{Fig:1}) to time $t$. The three parameters
($D$, $\tau$ and $y_0$) are determined by a nonlinear least-square
fitting algorithm. Besides, as the geomagnetic activity ($aa$
index) often lags behind solar flares by several months, the
relationships between $aa$ and the flare parameters are also
analyzed by the same model.

\subsection{Relationship between $R_\mathrm{z}$-$I$-$aa$} \label{subsec:I}

 \begin{figure}[!tb]
 \includegraphics[width=\columnwidth]{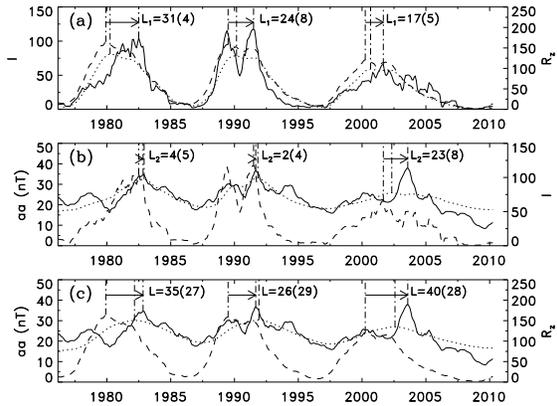}
 \caption{ (a)  $I$ (solid), $R_\mathrm{z}$ (dashed), and the reconstructed series  ($I_\mathrm{f}$, dotted)
     by Equation~(\ref{Eq:integral31}).
     The correlation coefficients of $I$ with $R_\mathrm{z}$ and $I_\mathrm{f}$
     are $r_0=0.87$ and $r_\mathrm{f}=0.88$, respectively.
     The lag times of $I$ ($I_\mathrm{f}$) to $R_\mathrm{z}$ at their peaks for Cycles 21-23 are $L_1=31,24,17$
     ($L_\mathrm{f1}=4,8,5$) months.
   (b) Similar results for the relationship between $aa$ (solid)
   and $I$ (dashed). The correlation coefficients of $aa$ with $I$ and
   the reconstructed series $aa_\mathrm{f}$ by Equation~(\ref{Eq:integral32})
   are $r_0=0.61$ and $r_\mathrm{f}=0.74$, respectively.
   The lag times of $aa$ ($aa_\mathrm{f}$) to $I$ at their peaks for Cycles 21-23 are $L_2=4,2,23$ ($L_\mathrm{f2}=5,4,8$) months.
     (c) For the relationship between $aa$ (solid)
     and $R_\mathrm{z}$ (dashed). The correlation coefficients of $aa$ with $R_\mathrm{z}$ and
     the reconstructed series $aa_\mathrm{f}$ by Equation~(\ref{Eq:integral30}) are $r_0=0.51$ and $r_\mathrm{f}=0.73$, respectively.
     The lag times of $aa$ ($aa_\mathrm{f}$) to $R_\mathrm{z}$ at their peaks for Cycles 21-23 are $L=35,26,40$
     ($L_\mathrm{f}=27,29,28$) months.
     }
 \label{Fig:2}
 \end{figure}

First, we analyze the relationship between $y=I$ and
$x=R_\mathrm{z}$ since March 1976 ($t_0$) with Equation
(\ref{Eq:model1}) in the form of
\begin{eqnarray}
    \label{Eq:integral31}
   \begin{array}{lrl}
    I(t)    &=& D_1\sum_{t'=t_0}^t R_\mathrm{z}(t')
            e^{-(t-t')/\tau_1}+I_0.
   \end{array}
\end{eqnarray}
Figure~\ref{Fig:2}(a) plots the reconstructed series
$I_\mathrm{f}$ (dotted) of $I$ (solid) from $R_\mathrm{z}$
(dashed) by Equation~(\ref{Eq:integral31}). Although the
correlation coefficient between $I$ and $R_\mathrm{z}$
($r_0=0.87$) has not been significantly improved by this model
($r_\mathrm{f}=0.88$), the lag times of $I$ to $R_\mathrm{z}$ at
their peaks (time differences between the peak timings) for Cycles
21-23 ($L_1=31,24,17$ with a mean $\overline{L}_1=24$ months) can
be predicted in part by Equation~(\ref{Eq:integral31}) as shown in
Fig.~\ref{Fig:2}(a) for the corresponding ones in brackets
($L_\mathrm{f1}=4,8,5$ with a mean $\overline{L}_\mathrm{f1}=6$).
It implies that the current flares are related to the accumulation
of solar magnetic energy in the past through a time decay factor.
Active magnetic structures may evolve from the photosphere to
upper chromosphere with different speeds and times
\citep{Lin95,Wheatland01}.

The relationship between $y=aa$ and $x=I$ can be fitted by
\begin{eqnarray}
    \label{Eq:integral32}
   \begin{array}{lrl}
    aa(t)    &=& D_2\sum_{t'=t_0}^t I(t') e^{-(t-t')/\tau_2}+aa'_0,
   \end{array}
\end{eqnarray}
as shown in Fig.~\ref{Fig:2}(b): $aa$ (solid), $I$ (dashed), and
the reconstructed series $aa_\mathrm{f}$ (dotted) by
Equation~(\ref{Eq:integral32}). One can see that $aa_\mathrm{f}$
reflects well the profile of $aa$. The correlation coefficient
between $aa$ and $aa_\mathrm{f}$ ($r_\mathrm{f}=0.74$) is higher
than that between $aa$ and $I$ ($r_0=0.61$). About half of the lag
times of $aa$ to $I$ at their peaks for Cycles 21-23 ($L_2=4,2,23$
with a mean $\overline{L}_2=10$) can be predicted by
Equation~(\ref{Eq:integral32}) as shown in Fig.~\ref{Fig:2}(b) for
the corresponding ones in brackets ($L_\mathrm{f2}=5,4,8$ with a
mean $\overline{L}_\mathrm{f2}=6$).

The relationship between $y=aa$ and $x=R_\mathrm{z}$ is analyzed
by using the following equation,
\begin{eqnarray}
    \label{Eq:integral30}
   \begin{array}{lrl}
    aa(t)       &=& D\sum_{t'=t_0}^t R_\mathrm{z}(t')
                e^{-(t-t')/\tau}+aa_0.
   \end{array}
\end{eqnarray}
Figure~\ref{Fig:2}(c) illustrates $aa$ (solid), $R_\mathrm{z}$
(dashed), and the reconstructed series $aa_\mathrm{f}$ (dotted) by
this equation. The correlation coefficient between $aa$ and
$aa_\mathrm{f}$ ($r_\mathrm{f}=0.73$) is much higher than that
between $aa$ and $R_\mathrm{z}$ ($r_0=0.51$). The lag times of
$aa$ to $R_\mathrm{z}$ at their peaks for Cycles 21-23
($L=35,26,40$ with a mean $\overline{L}=34$) can be well predicted
by Equation~(\ref{Eq:integral30}) as shown in Fig.~\ref{Fig:2}(c)
for the corresponding ones in brackets ($L_\mathrm{f}=27,29,28$
with a mean $\overline{L}_\mathrm{f}=28$). The above results are
listed in Table~\ref{Tab:tab1}, in which $\sigma$ refers to the
standard deviation, the last column indicates the relevant
averages of fitted/observed lag times at the corresponding peaks
over Cycles 21-23 ($\overline{L}_\mathrm{f}/\overline{L}$), and
the last three rows represent the relevant averages of the
parameters for the relationships between $P$-$R_\mathrm{z}$,
$aa$-$P$, and $aa$-$R_\mathrm{z}$, respectively, where $P=I, T, Q,
F$.

\begin{table*}[!tb]
 \small 
 \tabcolsep 1.2mm
 \caption{Fitted Results of the Integral Response Model for the Flare Parameter: $P=I, T, Q, F$.}
  \label{Tab:tab1}
 \begin{tabular}{llc|crrllrcccc}
 \tableline  
 $y$& $x$& $t_0$ &$D$& $\tau$& $y_0$& $r_0$& $r_\mathrm{f}$&
$\sigma$&
 $L_\mathrm{f}/L$(21)& $L_\mathrm{f}/L$(22)& $L_\mathrm{f}/L$(23)&
 $\overline{L}_\mathrm{f}/\overline{L}^{\ \mathrm{a}}$\\
  \tableline
 $I$&           $R_\mathrm{z}$&   Mar. 1976&   $9.88\times10^{-2}$   &4.9   &$-2.2$ &0.87 &0.88 &14.1   &4/31  &8/24  &5/17   &6/24  \\
 $aa$&          $I$           &   Mar. 1976&   $1.12\times10^{-2}$   &16.7  &16.7   &0.61 &0.74 &3.9    &5/4   &4/2   &8/23   &6/10  \\
 $aa$&          $R_\mathrm{z}$&   Mar. 1976&   $4.79\times10^{-3}$   &24.5  &15.1   &0.51 &0.73 &4.0    &27/35 &29/26 &28/40  &28/34 \\
  \tableline
 $T$&           $R_\mathrm{z}$&   Mar. 1976&   $9.24\times10^{-2}$   &14.2  &41.7   &0.79 &0.89 &29.6   &22/21 &27/21 &25/22  &25/21 \\
 $aa$&           $T$          &   Mar. 1976&   $7.42\times10^{-3}$   & 8.8  &14.0   &0.66 &0.74 &4.0    &7/14  &6/5   &2/18   &5/12  \\
 $aa$&          $R_\mathrm{z}$&   Mar. 1976&   $4.79\times10^{-3}$   &24.5  &15.1   &0.51 &0.73 &4.0    &27/35 &29/26 &28/40  &28/34 \\
  \tableline
 $Q$&           $R_\mathrm{z}$&   Jul. 1966&        0.094             &0.5   &$-1.3$ &0.90 &0.90 &2.3    &0/15  &0/-1  &0/0    &0/5  \\
 $aa$&          $Q$&              Jul. 1966&    $2.78\times10^{-2}$   &27.7  &19.1   &0.37 &0.58 &3.9    &18/20 &29/27 &24/40  &24/29 \\
 $aa$&          $R_\mathrm{z}$&   Jul. 1966&    $3.33\times10^{-3}$   &36.9  &15.0   &0.32 &0.64 &3.7    &31/35 &31/26 &30/40  &31/34 \\
  \tableline
 $F$&           $R_\mathrm{z}$&    Jul. 1997&  $3.50\times10^{-5}$   &13.6  &0.001  &0.79 &0.89 &0.009  & ---  & ---  &25/17  &25/17\\
 $aa$&          $F$&               Jul. 1997&  229.4                 &0.3   &14.0   &0.75 &0.75 &4.4    & ---  & ---  &0/23   &0/23\\
 $aa$&          $R_\mathrm{z}$&    Jul. 1997&   $8.32\times10^{-3}$  &18.6  &12.3   &0.62 &0.79 &4.0    & ---  & ---  &26/40  &26/40\\
  \tableline\tableline
\multicolumn{2}{l}{Av.($P$-$R_\mathrm{z}$)$^{\mathrm{b}}$}  &   &    &8.3   &       &0.84 &0.89 &       &9/22  &12/15 &14/14 &14/17 \\
 \multicolumn{2}{l}{Av.($aa$-$P$)}                         &   &    &13.4   &       &0.60 &0.70 &       &10/13 &3/11  &9/26 &9/19 \\
 \multicolumn{2}{l}{Av.($aa$-$R_\mathrm{z}$)}              &   &    &26.1   &       &0.49 &0.72 &       &28/35 &30/26 &28/40 &28/36 \\
 \tableline 
 \end{tabular}
 \begin{list}{}{}
  \item [$ ^{\mathrm{a}}$] Average over Cycles 21-23.
  \item [$ ^{\mathrm{b}}$] Average of the corresponding parameters for the relationships between
  $P$ ($=I, T, Q, F$) and $R_\mathrm{z}$.
  \end{list}
 \end{table*}

\subsection{Relationship between $R_\mathrm{z}$-$T$-$aa$} \label{subsec:T}

 \begin{figure}[!tb]
 \includegraphics[width=\columnwidth]{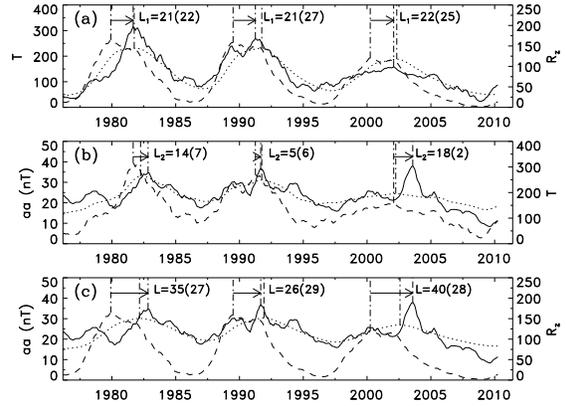}
 \caption{Similar to Fig.~\ref{Fig:2} for the relationship
between $R_\mathrm{z}$-$T$-$aa$}
 \label{Fig:3}
 \end{figure}

The relationship between $R_\mathrm{z}$-$T$-$aa$ since March 1976
($t_0$) can also be analyzed by the technique in the previous
section, with the results shown in Fig.~\ref{Fig:3} and
Table~\ref{Tab:tab1}.  The following can be noted.
\begin{enumerate}
  \item[(i)] The correlation coefficient of $T$ with the
reconstructed series $T_\mathrm{f}$ ($r_\mathrm{f}=0.89$) from
$R_\mathrm{z}$ by Equation~(\ref{Eq:model1}) is higher than that
of $T$ with $R_\mathrm{z}$ ($r_0=0.79$).
  \item[(ii)]  The correlation coefficient of $aa$
with the reconstructed series $aa_\mathrm{f}$
($r_\mathrm{f}=0.74$) from $T$ by Equation~(\ref{Eq:model1}) is
higher than that of $aa$ with $T$ ($r_0=0.66$).
  \item[(iii)] The lag times of
   $T$ to $R_\mathrm{z}$ ($\overline{L}_\mathrm{f1}/\overline{L}_1=25/21$),
   $aa$ to $T$ ($\overline{L}_\mathrm{f2}/\overline{L}_2=5/12$),
  and $aa$ to $R_\mathrm{z}$ ($\overline{L}_\mathrm{f}/\overline{L}=28/34$)
  at their peaks for Cycles 21-23 can be approximately predicted by the model.
\end{enumerate}

\subsection{Relationship between $R_\mathrm{z}$-$Q$-$aa$} \label{subsec:Q}

 \begin{figure}[!tb]
 \includegraphics[width=\columnwidth]{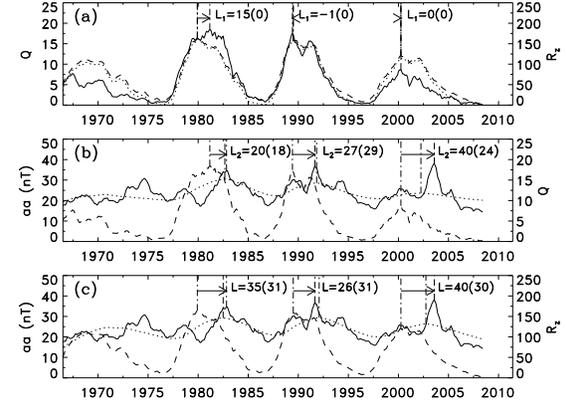}
 \caption{Similar to Fig.~\ref{Fig:2} for the relationship
between $R_\mathrm{z}$-$Q$-$aa$.}
 \label{Fig:4}
 \end{figure}

Figure~\ref{Fig:4} illustrates the relationship between
$R_\mathrm{z}$-$Q$-$aa$ since July 1966 ($t_0$) by using the
technique in Section~\ref{subsec:I}. One can note the following.
\begin{enumerate}
  \item[(i)] The correlation coefficient of $Q$ with the
reconstructed series $Q_\mathrm{f}$ ($r_\mathrm{f}=0.90$) from
$R_\mathrm{z}$ by Equation~(\ref{Eq:model1}) has not improved in
comparison to that of $Q$ with $R_\mathrm{z}$ ($r_0=0.90$),
implying that $Q$ and $R_\mathrm{z}$ peak nearly at the same time
\citep{Kleczek52}.
  \item[(ii)]  The correlation coefficient of $aa$
with the reconstructed series $aa_\mathrm{f}$
($r_\mathrm{f}=0.58$) from $Q$ by Equation~(\ref{Eq:model1}) is
much higher than that of $aa$ with $Q$ ($r_0=0.37$).
  \item[(iii)]  The correlation coefficient of $aa$
with the reconstructed series $aa_\mathrm{f}$
($r_\mathrm{f}=0.64$) from $R_\mathrm{z}$ by
Equation~(\ref{Eq:model1}) is much higher than that of $aa$ with
$R_\mathrm{z}$ ($r_0=0.32$).
  \item[(iv)]
  The lag times of
   $Q$ to $R_\mathrm{z}$ ($\overline{L}_\mathrm{f1}/\overline{L}_1=0/5$),
   $aa$ to $Q$ ($\overline{L}_\mathrm{f2}/\overline{L}_2=24/29$),
  and $aa$ to $R_\mathrm{z}$ ($\overline{L}_\mathrm{f}/\overline{L}=31/34$)
 at their peaks for Cycles 21-23 can be predicted in part by Equation
(\ref{Eq:model1}).
\end{enumerate}

\subsection{Relationship between $R_\mathrm{z}$-$F$-$aa$} \label{subsec:F}

\begin{figure}[!tb]
 \includegraphics[width=\columnwidth]{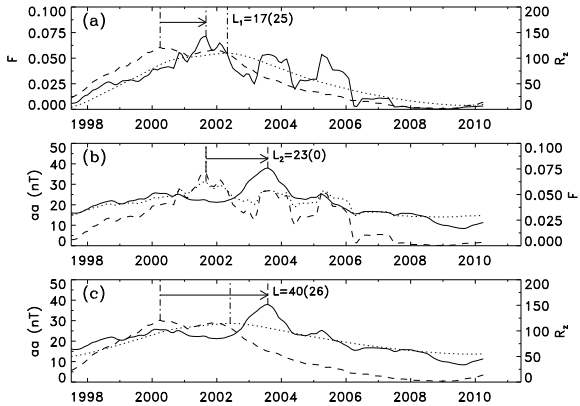}
 \caption{Similar to Fig.~\ref{Fig:2} for the relationship
between $R_\mathrm{z}$-$F$-$aa$.}
 \label{Fig:5}
 \end{figure}

Figure~\ref{Fig:5} shows the relationship between
$R_\mathrm{z}$-$F$-$aa$ since July 1997 ($t_0$) using the
technique in Section~\ref{subsec:I}. One sees the following.
\begin{enumerate}
  \item[(i)] The correlation coefficient of $F$ with the
reconstructed series $F_\mathrm{f}$ ($r_\mathrm{f}=0.89$) from
$R_\mathrm{z}$ by Equation~(\ref{Eq:model1}) is higher than that
of $F$ with $R_\mathrm{z}$ ($r_0=0.79$).
  \item[(ii)]  The correlation coefficient of $aa$
with the reconstructed series $aa_\mathrm{f}$
($r_\mathrm{f}=0.75$) from $F$ by Equation~(\ref{Eq:model1}) is
equal to that of $aa$ with $F$ ($r_0=0.75$).
  \item[(iii)]  The correlation coefficient of $aa$
with the reconstructed series $aa_\mathrm{f}$
($r_\mathrm{f}=0.79$) from $R_\mathrm{z}$ by
Equation~(\ref{Eq:model1}) is much higher than that of $aa$ with
$R_\mathrm{z}$ ($r_0=0.62$).
  \item[(iv)] The lag times of
   $F$ to $R_\mathrm{z}$ ($L_\mathrm{f1}/L_1=25/17$),
  and $aa$ to $R_\mathrm{z}$ ($L_\mathrm{f}/L=26/40$)
  at their peaks for Cycle 23 can be predicted in part by Equation (\ref{Eq:model1}).
  While the lag time of
   $aa$ to $F$ ($L_\mathrm{f2}/L_2=0/23$)
  at their peaks for Cycle 23 has not been predicted by Equation (\ref{Eq:model1}) due to the
  great fluctuations in both $aa$ and $F$.
\end{enumerate}

These results imply that solar flares depend not only on the
present but also on past solar activities ($R_\mathrm{z}$),
reflecting the long-term evolution characteristics of solar
magnetic field structures (energy) evolving from the photosphere
to upper chromosphere \citep{Donnelly87,Zhang07,Lin09}. The
correlations between solar flare parameters and $R_\mathrm{z}$ are
not simply due to the time shifts \citep{Bachmann94}. Solar flares
may play a role for the formation of geomagnetic activity ($aa$)
from solar (magnetic field) activity ($R_\mathrm{z}$), although
the processes are not completely clear \citep{Cliver02}.

\section{Discussions and Conclusions}
\label{sec:Discussions}

In this study, we investigated the relationships between the solar
flare parameters ($P=I$, $T$, $Q$ and $F$) and sunspot activity
($R_\mathrm{z}$), and between geomagnetic activity ($aa$) and the
flare parameters via the integral response
model~(Equation\,(\ref{Eq:model1})). The results indicate that (i)
the correlation coefficients between the flare parameters and
$R_\mathrm{z}$ have increased about 6\% from
$\overline{r}_\mathrm{0}=0.84$ to $\overline{r}_\mathrm{f}= 0.89$
on average when using Equation~(\ref{Eq:model1}) and the time
delays at their peaks for Cycles 21-23 can be well predicted by
this model, $\overline{L}_\mathrm{f1}/\overline{L}_\mathrm{1}=
14/17=82\%$; (ii) the correlation coefficients between $aa$ and
the flare parameters have increased about 17\% from
$\overline{r}_\mathrm{0}=0.60$ to $\overline{r}_\mathrm{f}= 0.70$
on average when using Equation~(\ref{Eq:model1}) and half of the
time delays at their peaks can be predicted by this model,
$\overline{L}_\mathrm{f2}/\overline{L}_\mathrm{2}= 9/19=47\%$; and
(iii) the correlation coefficient between $aa$ and $R_\mathrm{z}$
has increased about 47\% from $\overline{r}_\mathrm{0}=0.49$ to
$\overline{r}_\mathrm{f}= 0.72$ on average when using
Equation~(\ref{Eq:model1}) and the time delays at their peaks can
be well predicted by this model,
$\overline{L}_\mathrm{f}/\overline{L}_\mathrm{}= 28/36=78\%$. This
model might be used to improve the solar flare prediction, which
should be studied in future.

\begin{figure}[!tb]
 \includegraphics[width=\columnwidth]{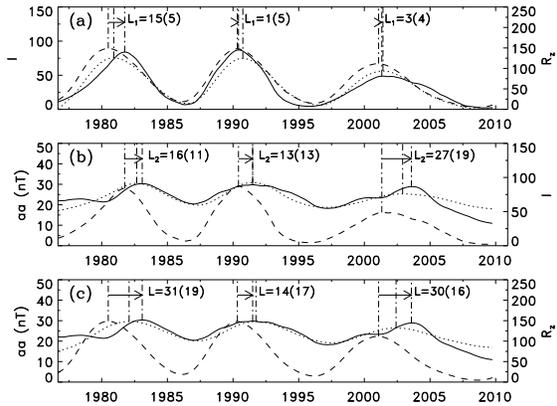}
 \caption{Same to Fig.~\ref{Fig:2} but using the cosine
filter (Equation~(\ref{Eq:cos})).}
 \label{Fig:6}
 \end{figure}

It is seen in Fig.~\ref{Fig:2}(a) and Table~\ref{Tab:tab1} that
the time delays between $I$ and $R_\mathrm{z}$ at their peaks for
Cycles 21-23 have not been well predicted by the model (6/24 =
25\%). This is due to the large fluctuations in the data. To
suppress further the fluctuations, we introduce a cosine filter
with the weights given by
\begin{equation}
  \label{Eq:cos}
        W_{\mathrm{C}}(\Delta t)=\frac{\pi}{4b}\cos(\frac{\pi\Delta
        t}{2b})
\end{equation}
for $b=24$ months. Since $\int_{-b}^b
\frac{\pi}{4b}\cos(\frac{\pi\Delta t}{2b})=1$, the weights
$W_{\mathrm{C}}(\Delta t)$ are normalized. Using the series
smoothed by this filter, we re-analyze the results in
Fig.~\ref{Fig:2}, as shown in Fig.~\ref{Fig:6}.

The time delays between $I$ and $R_\mathrm{z}$ at their peaks are
now better predicted, $L_\mathrm{1f}/L_1=5/15$, 5/1 and 4/3 for
Cycles 21-23 (Fig.~\ref{Fig:6}(a)), respectively, with a mean of
$\overline{L}_\mathrm{1f}/\overline{L}_\mathrm{1}=$ 4.7/6.3 = 74\%
which is much higher than the original one (6/24 = 25\%). The time
delays between $aa$ and $I$ at their peaks are also better
predicted, $L_\mathrm{2f}/L_2=11/16$, 13/13 and 19/27 for Cycles
21-23 (Fig.~\ref{Fig:6}(b)), respectively, with a mean of
$\overline{L}_\mathrm{2f}/\overline{L}_\mathrm{2}=$ 14.3/18.7 =
77\% which is higher than the original one (6/10 = 60\%). In
Fig.~\ref{Fig:6}(c), the time delays between $aa$ and
$R_\mathrm{z}$ at their peaks are predicted as
$L_\mathrm{f}/L=19/31$, 17/14 and 16/30 for Cycles 21-23,
respectively, with a mean of
$\overline{L}_\mathrm{f}/\overline{L}_\mathrm{}=$ 17.3/25 = 69\%
which is smaller than the original one (28/34 = 82\%) due to the
great lag time of $aa$ to $R_\mathrm{z}$ (about 30 months) and
other sources of $aa$.

In Equation~(\ref{Eq:model1}), the output $y$ depends on the past
values of input $x$ ($\tau>0$) rather than only the current value
($\tau=0$). The stronger the input ($x$), the more it contributes
to the output ($y$), and the longer the lag time of $y$ to $x$
\citep{Du11c}. Therefore, solar flares are related to the
accumulation of solar magnetic energy in the past rather than the
simple time shifts of occurrences \citep{Bachmann94}. The average
response time scale of flare parameters to $R_\mathrm{z}$ in this
model ($\overline{\tau}=8$) is close to the coronal response time
($\sim10$ months) derived from a model for dynamical energy
balance in the flaring solar corona
\citep{Wheatland01,Litvinenko04}.

There are various types of active regions in a solar cycle. Small
active regions of simple magnetic structure are short-lived and
produce minor solar flares, while large active regions of complex
magnetic structure are long-lived and produce major solar flares
(and hard X-ray flares). It is seen in Fig.~\ref{Fig:1}(a) that
$T$ is well correlated with $I$ ($r=0.89$), with the regression
equation given by
\begin{equation}
  \label{Eq:TI}
        T=67.9\pm 2.2 +(1.94\pm0.05)I.
\end{equation}
According to the above discussions, minor (low-energy) solar
flares lag behind the input $R_\mathrm{z}$ shorter times with
shorter durations while major solar flares lag behind
$R_\mathrm{z}$ longer times with longer durations. Therefore, (i)
the time delays between flare activities and sunspot activity come
mainly from the major flares rather than the weak ones; (ii) major
flares tend to have longer durations and may occur until quite
late in the decay phase of a solar cycle \citep{Temmer03,Tan11};
and (iii) the upper chromospheric activity indices
\citep{Donnelly87,Bachmann94} and the solar flares
\citep{Wheatland01,Temmer03} tend to lag behind the sunspot number
by several months 
in a hierarchy manner \citep{Bachmann94}.

Although it is unclear how solar flares affect geomagnetic
activities \citep{Gosling93,Cliver02}, it is apparent that
geomagnetic activities are well correlated with the solar flares.
For example, $aa$ is well correlated with $F$ ($r_0=0.75$). As
flares are unable to travel to 1 A.U., streams of matter emanating
from large flares were considered as the prime cause of
geomagnetic storms \citep{Hale31,Chapman50,Pudovkin77}. However,
\citet{Gosling93} argued that CMEs, not flares, were the critical
element for large geomagnetic storms, interplanetary shocks, and
major solar energetic particle (SEP) events. In fact, solar flares
may affect geomagnetic activities via different processes related
to the flare brightening, erupting, particle ejections, and other
unknown effects \citep{Cliver02}. Therefore, the relationships
between geomagnetic activity ($aa$) and solar flares can also be
well described by Equation~(\ref{Eq:model1}). Since geomagnetic
activity ($aa$) can be resulted from various activity phenomena
\citep{Legrand89,Tsurutani95}, the geomagnetic activity is the
integral of the effects of all these phenomena, including solar
winds, CMEs, solar flares and others. The lag time of $aa$ to
solar flare has not been well predicted by the model (9/19) due to
the additional effects of other activities. While the lag times of
both solar flare and $aa$ to $R_\mathrm{z}$ at their peaks have
been well predicted by the model (14/17, 28/36) because the solar
magnetic field activity is the main source of them.

The main conclusions can be drawn as follows,
\begin{enumerate}
  \item[(i)] The relationships between the flare parameters ($P=I, T, Q, F$) and sunspot
  activity ($R_\mathrm{z}$) can be well described by an integral response
  model ($r=0.89$)
  with a mean response time scale of about eight months.
  The time
  delays between the flare parameters and $R_\mathrm{z}$ at their peaks can be well predicted by
  this model ($82\%$).
  \item[(ii)] The relationships between geomagnetic activity ($aa$) and
  the flare parameters can be better described by this model ($r=0.70$)
  with a mean response time scale of thirteen months
  than by a linear dependence ($r=0.60$). The time
  delays between $aa$ and the flare parameters at their peaks can be predicted in half by
  this model ($47\%$).
  \item[(iii)] The relationship between $aa$ and
  $R_\mathrm{z}$ can be much better described by this model ($r=0.72$)
  with a mean response time scale of about twenty-six months
  than by a linear dependence ($r=0.49$). The time
  delay between $aa$ and $R_\mathrm{z}$ at their peaks can be predicted in part by
  this model ($78\%$).
\end{enumerate}

\section*{Acknowledgments}
The authors are grateful to the anonymous referee for constructive
comments. This work is supported by National Natural Science
Foundation of China (NSFC) through grants 10973020, 40890161 and
10733020, and National Basic Research Program of China (973
Program) through grant No. 2011CB811406.


\end{document}